\documentclass[french,english,reprint,amssymb,amsmath,cha,onecolumn,showpacs,showkeys,aps]{revtex4}
\usepackage[T1]{fontenc}
\usepackage[latin9]{inputenc}
\usepackage[active]{srcltx}
\usepackage{color}
\usepackage{babel}
\makeatletter
\addto\extrasfrench{%
   \providecommand{\fg}{\ifdim\lastskip>\z@\unskip\fi~\frqq}%
}

\makeatother
\usepackage{float}
\usepackage{amssymb}
\usepackage{graphicx}
\usepackage{esint}
\usepackage[unicode=true,pdfusetitle,
 bookmarks=true,bookmarksnumbered=false,bookmarksopen=false,
 breaklinks=false,pdfborder={0 0 1},backref=section,colorlinks=true]
 {hyperref}
\hypersetup{
 linkcolor=blue}
\usepackage{breakurl}

\makeatletter
\@ifundefined{textcolor}{}
{%
 \definecolor{BLACK}{gray}{0}
 \definecolor{WHITE}{gray}{1}
 \definecolor{RED}{rgb}{1,0,0}
 \definecolor{GREEN}{rgb}{0,1,0}
 \definecolor{BLUE}{rgb}{0,0,1}
 \definecolor{CYAN}{cmyk}{1,0,0,0}
 \definecolor{MAGENTA}{cmyk}{0,1,0,0}
 \definecolor{YELLOW}{cmyk}{0,0,1,0}
}

\usepackage{docs}
\usepackage{bm}
\expandafter\ifx\csname package@font\endcsname\relax\else
 \expandafter\expandafter
 \expandafter\expandafter\expandafter
 \expandafter{\csname package@font\endcsname}%
\fi
\@ifundefined{showcaptionsetup}{}{%
 \PassOptionsToPackage{caption=false}{subfig}}
\usepackage{subfig}
\makeatother

\begin{document}

\title{The one-dimensional thermal properties for the relativistic harmonic
oscillators}

\author{Abdelamelk Boumali}

\email{boumali.abdelmalek@gmail.com}

\affiliation{Laboratoire de Physique Appliquée et Théorique, \\
Université de Tébessa, 12000, W. Tébessa, Algeria.}

\date{\today}
\begin{abstract}
In this paper, we want to improved the calculations of the thermodynamic
quantities of the relativistic Harmonic oscillator using the Hurwitz
zeta function. The comparison of our results with those obtained by
a method based on the Euler-MacLaurin approach has been made. 
\end{abstract}

\keywords{Klein-Gordon oscillator;Dirac oscillator; Euler-MacLaurin formula;
Hurwitz zeta function}

\maketitle

\section{Introduction}

The relativistic harmonic oscillator is one of the most important
quantum system, as it is one of the very few that can be solved exactly.

The Dirac relativistic oscillator (DO) interaction is an important
potential both for theory and application. It was for the first time
studied by Ito et al \cite{1}. They considered a Dirac equation in
which the momentum $\vec{p}$ is replaced by $\vec{p}-im\beta\omega\vec{r}$,
with $\vec{r}$ being the position vector, $m$ the mass of particle,
and $\omega$ the frequency of the oscillator. The interest in the
problem was revived by Moshinsky and Szczepaniak \cite{2}, who gave
it the name of Dirac oscillator (DO) because, in the non-relativistic
limit, it becomes a harmonic oscillator with a very strong spin-orbit
coupling term. Physically, it can be shown that the (DO) interaction
is a physical system, which can be interpreted as the interaction
of the anomalous magnetic moment with a linear electric field \cite{3,4}.
The electromagnetic potential associated with the DO has been found
by Benitez et al\cite{5}. The Dirac oscillator has attracted a lot
of interest both because it provides one of the examples of the Dirac's
equation exact solvability and because of its numerous physical applications
\cite{6,7,8,9}. Fortunately Franco-Villafane et al \cite{10}, in
order to vibrate this oscillator, exposed the proposal of the first
experimental microwave realization of the one-dimensional (DO). 

The thermal properties of the one-dimensional Dirac equation in a
Dirac oscillator interaction was at first considered by Pacheco et
al \cite{11} . The authors have been calculated the all thermal quantities
of the oscillator by using the Euler-MacLaurin formula. Although this
method allows to obtain the all thermal properties of the system,
the expansion of the partition function using it could be valid only
for higher temperatures regime, but not otherwise. Also, the partition
function at $T=0\mbox{K}$ reveals a total divergence (see Annexe
A). Encouraged by the experimental realization of a Dirac oscillator,
we are interested in: (i) to improve the calculations of the thermodynamics
properties for the relativistic harmonic oscillators in all range
of temperatures, and (ii) to remove the divergence appears in the
partition function at $T=0\mbox{K}$. Both objectives can be achieved
by using a method based on the zeta function \cite{12,13}. This method
has been used by \cite{14} with the aim of calculating the partition
function in the case of the graphene. We note here that the zeta function
has been applied successfully in different areas of physics, and the
examples vary from ordinary quantum and statistical mechanics to quantum
field theory (see for example \cite{15}).

Thus, the main goal of this paper is the improvement of the calculations
of all thermal quantities of the one-dimensional Dirac oscillator.
This work is organized as follows: In section. 2, we review the solutions
of both Dirac and Klein-Gordon oscillators in one dimension. Section.
3 is devoted to our numerical results and discussions. Finally, Section.
4 will be a conclusion.

\section{Review of the solutions of both Dirac and Klein-Gordon oscillators
in one dimension}

\subsection{one-dimensional Klein-Gordon oscillator}

The free Klein-Gordon oscillator is written by
\begin{equation}
\left(p_{x}^{2}-\frac{E^{2}-m_{0}^{2}c^{4}}{c^{2}}\right)\phi=0.\label{eq:1}
\end{equation}
In the presence of the interaction of the type of Dirac oscillator,
it becomes
\begin{equation}
\left[c^{2}\left(p_{x}+im_{0}\omega x\right)\cdot\left(p_{x}-im_{0}\omega x\right)-E^{2}+m_{0}^{2}\right]\phi\left(x\right)=0,\label{eq:2}
\end{equation}
or
\begin{equation}
\left(\frac{p_{x}^{2}}{2m_{0}}+\frac{m_{0}\omega{}^{2}}{2}x^{2}\right)\phi\left(x\right)=\left(\frac{m_{0}c^{2}\hbar\omega+E^{2}-m_{0}^{2}c^{4}}{2m_{0}c^{2}}\right)\phi\left(x\right)\equiv\tilde{E}\phi\left(x\right),\label{eq:3}
\end{equation}
with
\begin{equation}
\tilde{E}=\frac{m_{0}c^{2}\hbar\omega+E^{2}-m_{0}^{2}c^{4}}{2m_{0}c^{2}}.\label{eq:4}
\end{equation}
The equation (\ref{eq:3}) is the standard equation of a harmonic
oscillator in $1D$. The energy levels are well known, and the solutions
are 
\begin{equation}
\epsilon_{n}=\pm m_{0}c^{2}\sqrt{1+2rn}\label{eq:5}
\end{equation}
with $r=\frac{\hbar\omega}{m_{0}c^{2}}$ being a parameter which controls
the non relativistic limit.

The eigenfunctions may be expressed in terms of Hermite Polynomial
of Degree $n$ as
\begin{equation}
\phi\left(x\right)=\mbox{N}_{\mbox{norm}}\left(\frac{m_{0}\omega}{\pi\hbar}\right)^{\frac{1}{4}}H\left(\sqrt{\frac{m_{0}\omega}{\hbar}}x\right)e^{-\frac{m\omega}{2\hbar}x^{2}}.\label{eq:6}
\end{equation}
where the functions $H$ is the so called Hermite polynomials, and
$\mbox{N}_{\mbox{norm}}$ is a normalizing factor \cite{16}.

\subsection{one-dimensional Dirac oscillator}

The one-dimensional Dirac oscillator is 
\begin{equation}
\left\{ c\cdot\alpha_{x}\cdot\left(p_{x}-im\omega\beta\cdot x\right)+\beta mc^{2}\right\} \psi_{D}=E\psi_{D},\label{eq:7}
\end{equation}
with $\psi_{D}=\left(\begin{array}{cc}
\psi_{1} & \psi_{2}\end{array}\right)^{T}$, $\alpha_{x}=\sigma_{x}$ and $\beta=\sigma_{z}$.

From Eq. (\ref{eq:7}), we get a set of coupled equations as follows:
\begin{equation}
\left(E-mc^{2}\right)\psi_{1}=c\left(p_{x}+im\omega x\right)\psi_{2},\label{eq:8}
\end{equation}
\begin{equation}
\left(E+mc^{2}\right)\psi_{2}=c\left(p_{x}-im\omega x\right)\psi_{1}.\label{eq:9}
\end{equation}
Using Eq. (\ref{eq:9}), we have
\begin{equation}
\psi_{2}\left(x\right)=\frac{c\left(p_{x}-im\omega x\right)}{E+mc^{2}}\psi_{1}\left(x\right).\label{eq:10}
\end{equation}
Putting Eq. (\ref{eq:10}) into (\ref{eq:8}), we get

\begin{equation}
\left[c^{2}\left(p_{x}+im\omega x\right)\left(p_{x}-im\omega x\right)-E^{2}+m^{2}\right]\psi_{1}\left(x\right)=0,\label{eq:11}
\end{equation}
or
\begin{equation}
\left(\frac{p_{x}^{2}}{2m}+\frac{m\omega^{2}}{2}x^{2}\right)\psi_{1}\left(x\right)=\left(\frac{\hbar\omega mc^{2}+E^{2}-m^{2}c^{4}}{2mc^{2}}\right)\psi_{1}\left(x\right)\equiv\tilde{E}\psi_{1}.\label{eq:12}
\end{equation}
The equation (\ref{eq:12}) is the standard equation of a harmonic
oscillator in $1D$. The energy levels are well-known, and are given
by 
\begin{equation}
\varepsilon_{n}=\pm mc^{2}\sqrt{1+2\bar{r}n}\label{eq:13}
\end{equation}
with $\bar{r}=\frac{\hbar\omega}{mc^{2}}$ is a parameter which controls
the non relativistic limit. The eigenfunctions may be expressed in
terms of Hermite Polynomial of Degree $n$ a
\begin{equation}
\psi_{1}\left(x\right)=\mbox{N}_{\mbox{norm}}^{'}\left(\frac{m\omega}{\pi\hbar}\right)^{\frac{1}{4}}H\left(\sqrt{\frac{m\omega}{\hbar}}x\right)e^{-\frac{m\omega}{2\hbar}x^{2}}.\label{eq:14}
\end{equation}
with $\mbox{N}_{\mbox{norm}}^{'}$ is a normalizing factor. The total
associated wave function is
\begin{equation}
\psi_{D}\left(x\right)=\mbox{N}_{\mbox{norm}}^{'}\left[\begin{array}{c}
1\\
\frac{c\left(p_{x}-im\omega x\right)}{E+mc^{2}}
\end{array}\right]H\left(\sqrt{\frac{m\omega}{\hbar}}x\right)e^{-\frac{m\omega}{2\hbar}x^{2}}.\label{eq:15}
\end{equation}

\section{Thermal properties of the relativistic harmonic oscillator}

Before we study the thermodynamic properties of both oscillators,
we can see that the form of the spectrum of energy (see Eqs. (\ref{eq:5})
and (\ref{eq:13})), for both cases, is the same. As consequently,
the numerical thermal quantities found, for both oscillators, are
similar . Thus, we focus, firstly, on the study of the thermal properties
of a Dirac oscillator, and then all results obtained can be extended
to the case of the one-dimensional Klein-Gordon oscillator.

\subsection{Methods}

In order to obtain all thermodynamic quantities of the relativistic
harmonic oscillator, we concentrate, at first, on the calculation
of the partition function $Z$. The last is defined by
\begin{equation}
Z=\sum_{n}e^{-\beta E_{n}}=\sum_{n}e^{-\frac{\sqrt{1+2rn}}{\tau}}\label{eq:16}
\end{equation}
With the following substitutions:
\begin{equation}
\alpha=\frac{1}{2r},~\gamma=\sqrt{2r},\label{eq:17}
\end{equation}
it becomes
\begin{equation}
Z=\sum_{n}e^{-\frac{\gamma}{\tau}\sqrt{\alpha+n}},\label{eq:18}
\end{equation}
with $\tau=\frac{k_{B}T}{mc^{2}}$ denotes the reduce temperature.

Using the formula \cite{12,13}
\begin{equation}
e^{-x}=\frac{1}{2\pi i}\int_{C}dsx^{-s}\Gamma\left(s\right),\label{eq:19}
\end{equation}
the sum in Eq. (\ref{eq:18}) is transformed into
\begin{equation}
\sum_{n}e^{-\frac{\gamma}{\tau}\sqrt{\alpha+n}}=\frac{1}{2\pi i}\int_{C}ds\left(\frac{\gamma}{\tau}\right)^{-s}\sum_{n}\left(n+\alpha\right)^{-\frac{s}{2}}\Gamma\left(s\right)=\frac{1}{2\pi i}\int_{C}ds\left(\frac{\gamma}{\tau}\right)^{-s}\zeta_{H}\left(\frac{s}{2},\alpha\right)\Gamma\left(s\right),\label{eq:20}
\end{equation}
with $x=\frac{\gamma}{\tau}\sqrt{\alpha+n}$, and $\Gamma\left(s\right)$
and $\zeta_{H}\left(\frac{s}{2},\alpha\right)$ are respectively the
Euler and Hurwitz zeta function. Applying the residues theorem, for
the two poles $s=0$ and $s=2$, the desired partition function is
written down in terms of the Hurwitz zeta function as follows:
\begin{equation}
Z\left(\tau\right)=\frac{\tau^{2}}{2r}+\zeta_{H}\left(0,\alpha\right).\label{eq:21}
\end{equation}
Now, using that 
\begin{equation}
\zeta_{H}\left(0,\alpha\right)=\frac{1}{2}-\alpha,\label{eq:22}
\end{equation}
the final partition function is transformed into 
\begin{equation}
Z\left(\tau\right)=\frac{\tau^{2}}{2r}+\frac{1}{2}-\frac{1}{2r}.\label{eq:23}
\end{equation}
From this definition, all thermal properties of both fermionic and
bosonic oscillators can be obtained.

\subsection{Numerical results and discussions}

\begin{figure}[H]
\includegraphics{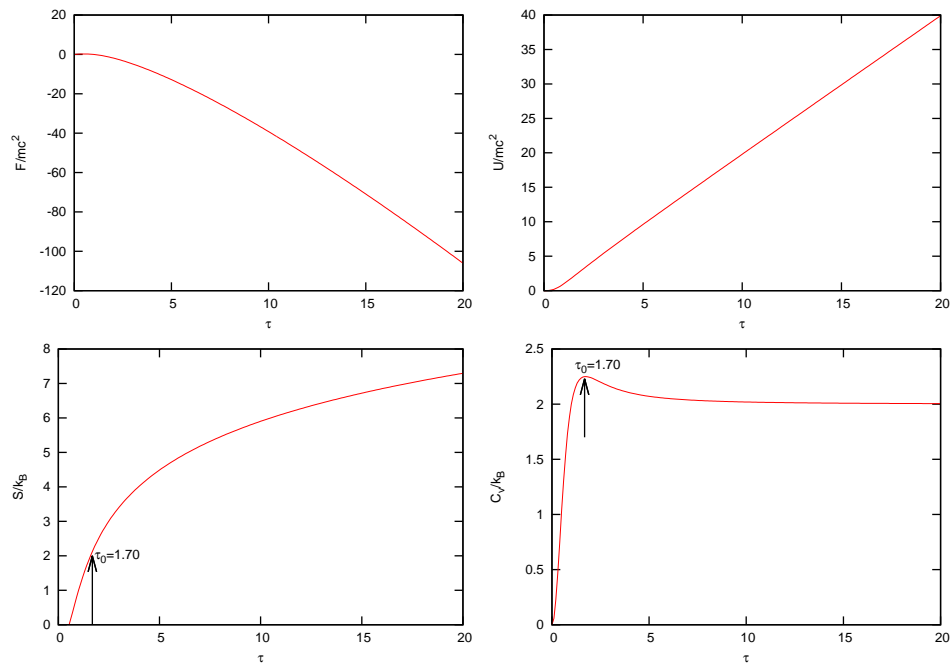}

\protect\caption{\label{fig: 1}One-dimensional thermal properties for both oscillators
in the case where $r=1$.}

\end{figure}

Fig. \ref{fig: 1} depicted the one-dimensional thermal properties
of both Dirac and Klein-Gordon oscillators. Following the figure,
all thermal quantities are plotted versus a reduced temperature $\tau$:
here we have taken $r=1$ which corresponds to the relativistic region
. From the curve of the numerical entropy function, no abrupt change,
around $\tau_{0}$, has been identified. This means that the curvature,
observed in the specific heat curve, does not exhibit or indicate
an existence of a phase transition around a $\tau_{0}$ temperature. 

\begin{figure}[H]
\subfloat[for the values of $r\geq0.5$]{\includegraphics{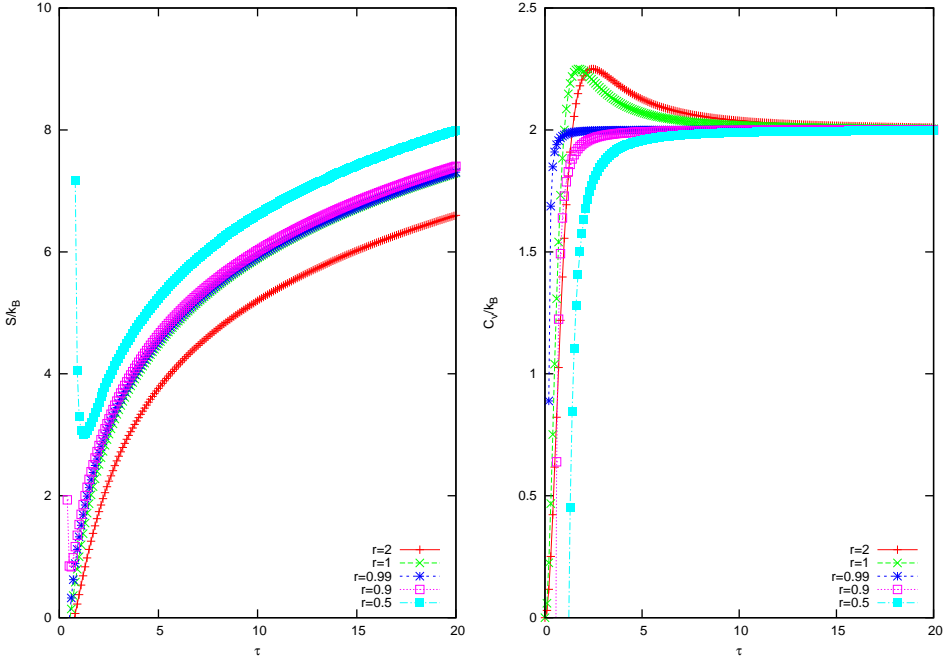}

}

\subfloat[for the values of $r<0.5$]{\includegraphics{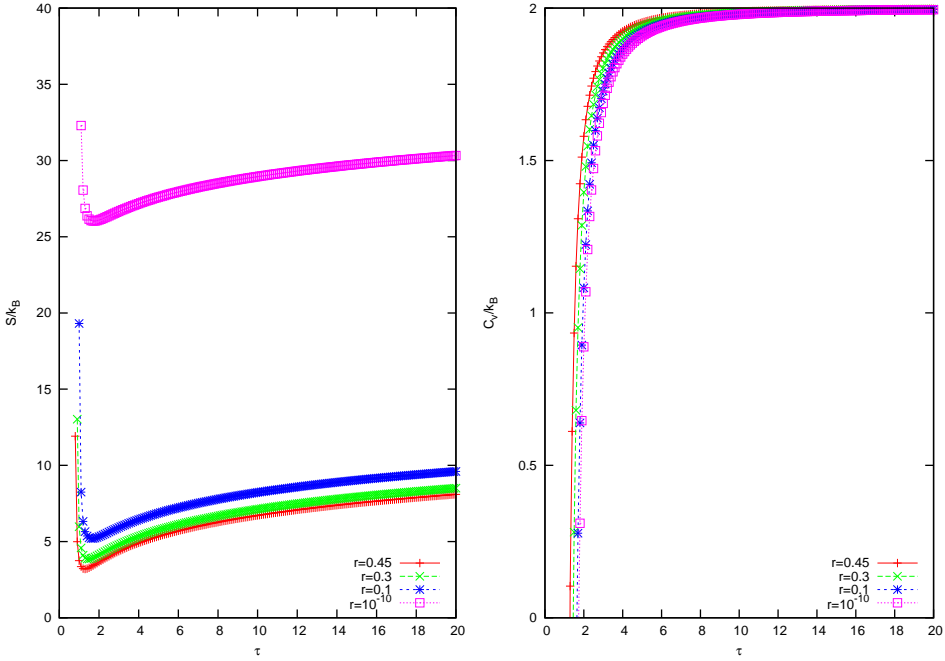}

}

\protect\caption{\label{fig:2}The entropy and specific heat as a function of a reduce
temperature $\tau$ for different values of the parameter $r$.}

\end{figure}

Now, according to the condition on $\alpha$ parameter which appears
in the zeta function (see Annexe B), we can distinguish two regions:
the first, defined by $r\geq0.5$, corresponds to the relativistic
regime, and the other, with $r<0.5$, represents the non-relativistic
regime. These observations, for both Dirac and Klein-Gordon oscillators,
are shown clearly in the Fig. \ref{fig:2}. Also, following the same
figure, three remarks can be made: 
\begin{itemize}
\item The $\tau_{0}$ reduce temperature increases for the values of $r>1$,
and disappears when $r<1$. 
\item When $r<0.5$ all curves coincide with the non-relativistic limit
($r=10^{-10}$).
\item For all values of $r$ , all curves of the specific heat coincide
with the limit $2k_{B}$.
\end{itemize}

Pacheco et al \cite{11} have been studied the thermal properties
of a Dirac oscillator in one dimension. They obtained all thermodynamics
quantities by using the Euler-MacLaurin approximation(see Annexe A).
The formalism used by \cite{11} is valid, only, for higher temperatures.
So, in order to cover all range of temperatures, we have employed
the Hurwitz zeta function method.

\begin{figure}
\includegraphics{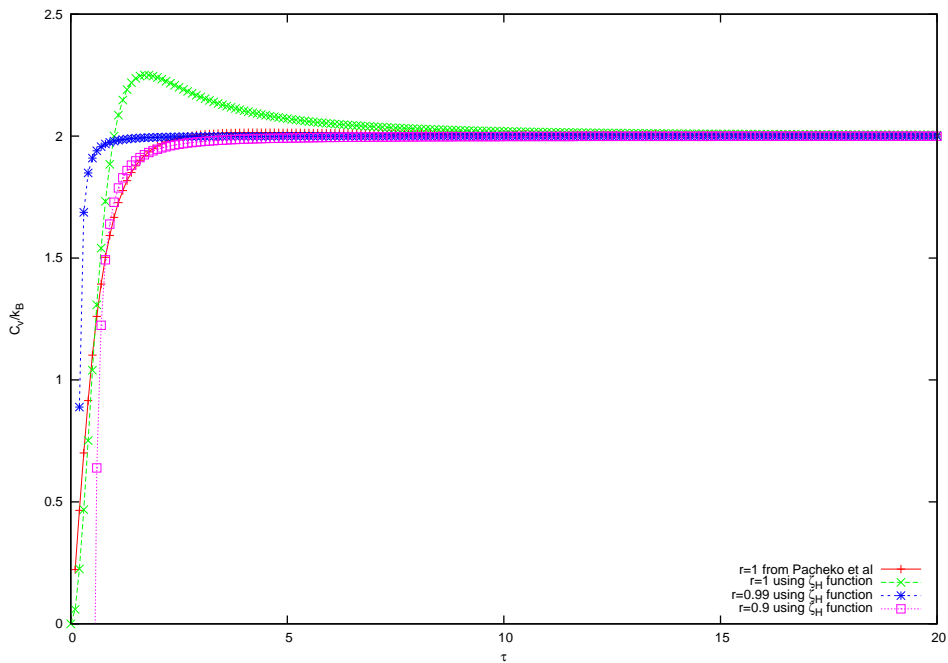}

\protect\caption{\label{fig:3}Comparison of our specific heat for different values
of $r$ with that obtained by using the Euler-MacLaurin formula for
a Dirac oscillator in one-dimension.}

\end{figure}

\begin{figure}
\includegraphics{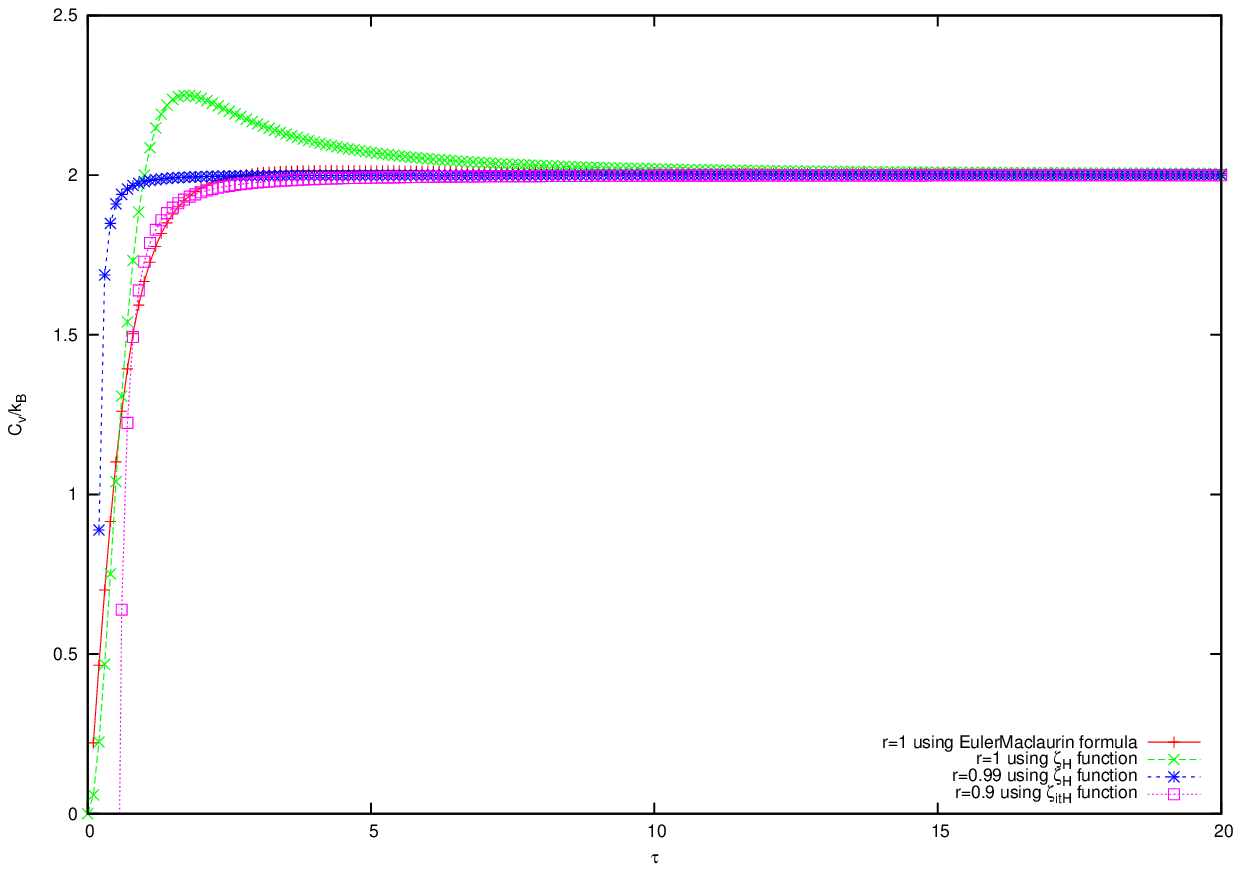}

\protect\caption{\label{fig:4}Comparison of our specific heat for different values
of $r$ with that obtained by using the Euler-MacLaurin formula for
a Klein-Gordon oscillator in one-dimension.}

\end{figure}

In the Fig. \ref{fig:3}, we are focused on the curves of the numerical
specific heat calculated for different values of the parameter $r$.
Then, for comparison with \cite{11}, we have inserted the numerical
calculation of the specific heat based on the Euler-MacLaurin approximation.
Thus, we can see that our results can be considered as an improvements
of the results obtained in \cite{11}. This consideration can be argued
as follows: (i) all thermal quantities obtained from the zeta function
method are valid in all range of temperatures, and (ii) the divergence
of the partition function, which appears in Euler-MacLaurin formula,
has been removed. All results obtained in this case can be extended
to the case of the Klein-Gordon oscillator (see Fig. \ref{fig:4}).

\begin{figure}[h]
\includegraphics[scale=0.5]{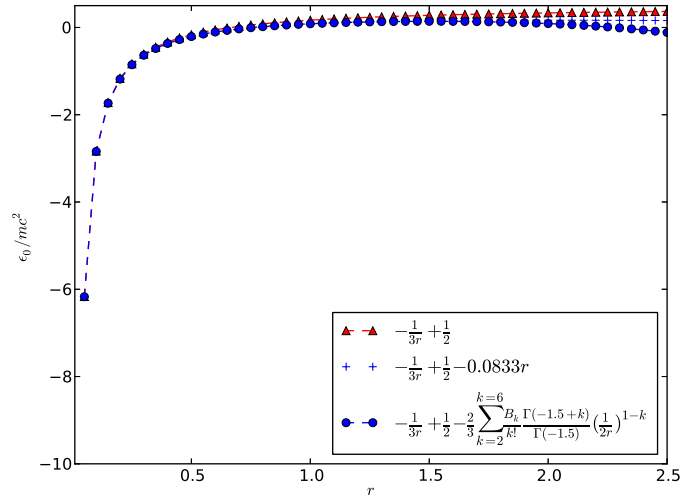}

\protect\caption{\label{fig:5}The reduce vacuum expectation value of the energy $\frac{\epsilon_{0}}{mc^{2}}$
versus a parameter $r$ around a relativistic region.}

\end{figure}

Finally, one may compute the vacuum expectation value of the energy
defined by \cite{15}
\begin{equation}
\epsilon_{0}=\lim_{s\rightarrow-1}\sum_{n=0}^{\infty}\left|\epsilon_{n}\right|^{-s}.\label{eq:24}
\end{equation}
From the spectrum of energy of a Dirac oscillator (Eq. (\ref{eq:13})),
the equation (\ref{eq:24}) can be expressed in terms of the $\zeta_{H}$
as follows:
\begin{equation}
\frac{\epsilon_{0}}{mc^{2}}=\sqrt{2r}\zeta_{H}\left(-\frac{1}{2},\frac{1}{2r}\right).\label{eq:25}
\end{equation}
Using the asymptotic series corresponding to the Hurwitz zeta function
(see Annexe B), Eq. (\ref{eq:25}) becomes
\begin{equation}
\frac{\epsilon_{0}}{mc^{2}}=-\frac{1}{3r}+\frac{1}{2}-\frac{2}{3}\sum_{k=2}^{\infty}\frac{B_{k}}{k!}\frac{\Gamma\left(-\frac{3}{2}+k\right)}{\Gamma\left(-\frac{3}{2}\right)}\left(\frac{1}{2r}\right)^{1-k}.\label{eq:26}
\end{equation}
In the Fig. (\ref{fig:5}), we show the $\frac{\epsilon_{0}}{mc^{2}}$
as a function of the parameter $r$. We can see that the vacuum expectation
value of the energy, which depend on the parameter $r$, can approximate
with
\begin{equation}
\frac{\epsilon_{0}}{mc^{2}}\approxeq-\frac{1}{3r}+\frac{1}{2}.\label{eq:27}
\end{equation}

\section{Conclusion}

In this work, we reviewed the relativistic harmonic oscillator for
both fermionic and bosonic massive particles in one dimension. The
statistical quantities of both Dirac and Klein-Gordon oscillators
were investigated by employing the zeta function method. The both
cases have been confronted with those obtained by using the Euler-MacLaurin
formula. The vacuum expectation value of the energy, for both oscillators,
has been estimated.

\appendix

\section{Euler-MacLaurin formula}

The partition function Z of the Dirac oscillator at finite temperature
$T$ is obtained through the Boltzmann factor \cite{7},
\begin{equation}
Z=\sum_{n=0}^{n}e^{-\beta\left(E_{n}-E_{0}\right)}=e^{\beta\sqrt{b}}\sum_{n=0}^{n}e^{-\beta\sqrt{an+b}},\label{eq:28}
\end{equation}
where $\beta=\frac{1}{k_{B}T}$ , $k_{B}$ is the Boltzmann constant,
$E_{0}$ is the ground state energy correspondent to $n=0$.

Before entering in the calculations, let us test the convergence of
the series of (\ref{eq:28}). For that, we apply the integral test
which shows that the series and the integral converge or diverge together.
So, from (\ref{eq:28}), we can see that the function $f(x)$, where
\begin{equation}
f\left(x\right)=e^{-\beta\sqrt{an+b}},\label{eq:29}
\end{equation}
is a decreasing positive function, and the integral
\begin{equation}
\int_{0}^{\infty}f\left(x\right)dx=\frac{2}{a\beta^{2}}e^{-\beta\sqrt{b}}\left(1+\beta\sqrt{b}\right),\label{eq:30}
\end{equation}
is convergent. This means that, according to the criterion of the
integral test, the numerical partition function Z converges. 

In order to evaluate this function, we use the Euler-MacLaurin formula
defined as follows 
\begin{equation}
\sum_{x=0}^{\infty}f\left(x\right)=\frac{1}{2}f\left(0\right)+\int_{0}^{\infty}f\left(x\right)dx-\sum_{p=1}^{\infty}\frac{1}{\left(2p\right)!}B_{2p}f^{\left(2p-1\right)}\left(0\right),\label{eq:31}
\end{equation}
Here, $B_{2p}$ are the Bernoulli numbers, $f^{(2p-1)}$ is the derivative
of order $\left(2p-2\right)$. In our case, we have taken $B_{2}=\frac{1}{6}$
and $B_{4}=-\frac{1}{30}$.

In our case, we have used
\begin{equation}
f^{\left(1\right)}\left(0\right)=\frac{a\beta}{2\sqrt{b}}e^{-\beta\sqrt{b}}\label{eq:32}
\end{equation}
\begin{equation}
f^{\left(3\right)}\left(0\right)=\left[\frac{-3\beta a^{3}}{8\left(b\right)^{5/2}}-\frac{3\beta^{2}a^{3}}{8\left(b\right)^{2}}-\frac{\beta^{3}a^{3}}{8\left(b\right)^{3/2}}\right]e^{-\beta\sqrt{b}}.\label{eq:33}
\end{equation}
Following Eqs. (\ref{eq:31}), (\ref{eq:32}) and (\ref{eq:33}),
the partition function can be cast into
\begin{equation}
Z=\frac{1}{2}+\frac{1}{2r}\tau+\frac{1}{2r}\tau^{2}+\left(\frac{r}{6}-\frac{r^{3}}{60}\right)\frac{1}{\tau}-\frac{1}{30}\frac{r^{3}}{\tau^{2}}-\frac{4}{45}\frac{r^{3}}{\tau^{3}}.\label{eq:34}
\end{equation}

\section{Some properties of zeta function}

The Riemann zeta function is defined by\cite{15}
\begin{equation}
\zeta\left(s\right)=\sum_{n=0}^{\infty}\frac{1}{n^{s}},~\mbox{with}~s\in\mathbb{C}.\label{eq:35}
\end{equation}
Nowadays the Riemann zeta function is just one member of a whole family
of zeta function's (Hurwitz, Epstein,Selberg). The most important
of them is the Hurwitz zeta function $\zeta_{H}$ given by
\begin{equation}
\zeta_{H}\left(s,\alpha\right)=\sum_{n=0}^{\infty}\frac{1}{\left(n+\alpha\right)^{s}},\label{eq:36}
\end{equation}
where $0<\alpha\leq1$, is a well-defined series when $\Re e\left(s\right)>1$,
and can be analytically continued to the whole complex plane with
one singularity, a simple pole with residue 1 at $s=1$.

An integral representation is
\begin{equation}
\zeta_{H}\left(s,\alpha\right)=\frac{1}{\Gamma\left(s\right)}\int_{0}^{\infty}dtt^{s-1}\frac{e^{-t\alpha}}{1-e^{-t}},~\Re\left(s\right)>1,~\Re\left(\alpha\right)>0.\label{eq:37}
\end{equation}
It can be shown that $\zeta_{H}\left(s,\alpha\right)$ has only one
singularity --namely a simple pole at $s=1$ with residue 1 and that
it can be analytically continued to the rest of the complex s-plane. 

Also, we can shown that $\zeta_{H}\left(s,\alpha\right)$ have the
following properties:
\begin{equation}
\zeta_{H}\left(0,\alpha\right)=\frac{1}{2}-\alpha,\label{eq:38}
\end{equation}
\begin{equation}
\zeta_{H}\left(-m,\alpha\right)=-\frac{B_{m+1}\left(\alpha\right)}{m+1},~m\in\mathbb{N},\label{eq:39}
\end{equation}
$B_{r}(a)$ being the Bernoulli polynomials. The asymptotic series
corresponding the Hurwitz zeta function is given by
\begin{equation}
\zeta_{H}\left(1+z,\alpha\right)=\frac{1}{z}\alpha^{-z}+\frac{1}{2}\alpha^{-1-z}+\frac{1}{z}\sum_{k=2}^{\infty}\frac{B_{k}}{k!}\frac{\Gamma\left(z+k\right)}{\Gamma\left(z\right)}\alpha^{-z-k},\label{eq:40}
\end{equation}
with $B_{k}$are Bernoulli's numbers.

\end{document}